\documentclass[useAMS,usenatbib,usegraphicx]{mn2e}

\def\kms{$\mathrm {km\,s}^{-1}$}
\def\mdot{$\mathrm {M_{\odot}\,yr}^{-1}$}
\def\msun{$\mathrm {M_{\odot}}$}
\def\density{$10^{-6}\ \mathrm {M_{\odot}\,pc}^{-3}$}

\def\oversim#1#2{\lower0.5pt\vbox{\baselineskip0pt \lineskip-0.5pt
     \ialign{$\mathsurround0pt #1\hfil##\hfil$\crcr#2\crcr\sim\crcr}}}
\title[Detached shells as tracers of AGB-ISM bow shocks]{Detached shells as tracers of AGB-ISM bow shocks}
\author[C. J. Wareing et al.]{C. J. Wareing$^{1}$\thanks{E-mail: cwareing@jb.man.ac.uk}, 
Albert A. Zijlstra$^{1}$, Angela K. Speck$^{2}$, T. J. O'Brien$^{1}$, 
\newauthor
Toshiya Ueta$^{3}$, M. Elitzur$^{4}$, R. D. Gehrz$^{5}$, F. Herwig$^{6}$, H. Izumiura$^{7}$, 
\newauthor
M. Matsuura$^{8}$, M. Meixner$^{9}$, R. E. Stencel$^{10}$, R. Szczerba$^{11}$\\
$^{1}$Jodrell Bank Centre for Astrophysics, University of Manchester, Oxford Road, Manchester, M13 9PL, UK\\
$^{2}$Department of Physics and Astronomy, 223 Physics Building, University of Missouri, Columbia, MO 65211, USA\\
$^{3}$NASA Ames Research Center/USRA SOFIA Office, Mail Stop 211-3, Moffett Field, CA 94035, USA\\
$^{4}$Physics \& Astronomy Department, University of Kentucky, Lexington, KY 40506, USA\\
$^{5}$Department of Astronomy, School of Physics and Astronomy, 116 Church Street, S.E., University of Minnesota,\\ \ Minneapolis, MN 55455, USA\\
$^{6}$Theoretical Astrophysics Group, LANL, Los Alamos, NM 87545, USA\\
$^{7}$Okayama Astrophysical Observatory, National Astronomical Observatory of Japan, Kamogata, Asakuchi,\\ \ Okayama 719-0232, Japan\\
$^{8}$Division of Optical and IR Astronomy, National Astronomical Observatory of Japan, Osawa 2-21-1, Mitaka,\\ \ Tokyo 181-8588, Japan\\
$^{9}$STScI, 3700 San Martin Dr., Baltimore, MD 21218, USA\\
$^{10}$Department of Physics and Astronomy, University of Denver, Denver, CO 80208, USA\\
$^{11}$N. Copernicus Astronomical Centre, Rabianska 8, 87-100, Torun, Poland}

\begin{document}

\date{submitted to MNRAS Letters}


\maketitle

\label{firstpage}

\begin{abstract}

New {\it Spitzer} imaging observations have
revealed the structure around the Mira variable star R Hya to be a 
one-sided parabolic arc 100 arcsec to the West stretching from North
to South. We successfully model R Hya and its surroundings in terms of
an interaction of the stellar wind from an asymptotic giant branch
(AGB) star with the interstellar medium (ISM) the star moves
through. Our three-dimensional hydrodynamic simulation reproduces the
structure as a bow shock into the oncoming ISM. We propose this as
another explanation of detached shells around such stars which should
be considered alongside current theories of internal origin. 
The simulation predicts the existence of a tail of
ram-pressure-stripped AGB material stretching downstream. Indications
for such a tail behind R Hya are seen in {\it IRAS} maps.

\end{abstract}


\section{Introduction}

Large detached shells have been observed around several asymptotic
giant branch (AGB) stars. They have been seen in IRAS images of dust
emission \citep{waters94,izumiura97}, CO line emission
\citep{olofsson96,schoier05}, and in a few cases [Na\,{\sc i}] and 
[K\,{\sc i}] emission as well as in the optical continuum
\citep{gonzalez01,gonzalez03}.  \cite{olofsson90} suggested that such
shells are the result of mass-loss variations and in particular, a
thermal pulse or He-flash. During a He-flash, an intense short-lived
mass ejection is driven by the star reaching a critical luminosity.
Thermal pulses are separated by phases of quiescent hydrogen burning
lasting $10^4$--$10^5$\,yr.  The stellar evolution tracks calculated
by \cite{vassiliadis93} confirmed that mass-loss fluctuations during
the thermal pulse cycle can lead to detached circumstellar shells.
Hydrodynamic simulations by \cite{steffen00}  showed
that a brief period of high mass loss can translate into a
geometrically thin shell expanding around the star. This has become
the standard explanation of detached dust shells around stars and
observations have been interpreted as such \citep{zijlstra92,speck00}.

A separate explanation for large detached shells is the interaction of
the AGB wind with the interstellar medium (ISM) \citep{young93}.  
\cite{zijlstra02a} used this for a giant ($\sim4$ pc
at D $\sim700$ pc) detached shell surrounding an M3 III AGB star. They
proposed that its AGB wind has been stopped by the surrounding ISM and
the swept-up `wall' is now expanding at the local sound
speed. Simulations by \cite{villaver03} and \cite{wareing06} confirmed
the viability of this mechanism.  It is likely that both mechanisms
occur, but whether the external mechanism of an ISM wall, or the
internal mechanism of long-term mass-loss variations is the dominant
cause of detached shells is not known. Sch{\"o}ier et al. (2005) found that the
derived masses of the shells increase and the expansion velocities
decrease with increasing radial distance from the star. They suggest
that the shell is sweeping
up surrounding material from an earlier mass loss phase. However, ISM
sweep-up could yield similar effects.  

A difference between the two
mechanisms is that the internal one will normally give a spherical
shell, while external mechanism will give a shape which depends on the
motion of the star through the ISM. This is a testable prediction,
if the proper motion of the star is known. Here we show that the detached shell
around the Mira variable R Hya is due to an ISM bow shock.

\section{Observations}


R Hya is one of the brightest Mira
variables on the sky.  \cite{hashimoto98} found evidence in IRAS data
for a detached shell 1--2 arcmin from the star. The star itself
coincides with an IRAS point source due to an inner dust shell. This
inner shell is also detached from the star, with an inner radius of
about 60 stellar radii. Since its discovery in AD 1662, the pulsation period has decreased
from 495 days to 385 days, attributed to non-linear pulsation or a
recent thermal pulse \citep{zijlstra02b}.  The inner detachment
suggests a mass-loss interruption occurred around AD 1800.  The
outer shell has a dynamical age of around 8000\,yr. This small time
difference causes problems if one assumes that both detached shells
are due to thermal pulses 
(Zijlstra et al. 2002).

The reported proper motion from the {\it Hipparcos} catalogue is 60.73
mas\,yr$^{-1}$ West and 11.01 mas\,yr$^{-1}$ North \citep{perryman97}. 
The radial velocity is $-10.4$ \kms\
\citep{wilson53}.  The distance to R Hya is uncertain, as discussed in
section 4 of 
Zijlstra et al. (2002). \cite{eggen1985}
argues for a distance of 165 pc based on a proper motion companion,
but the data this is based on appears to be unpublished. The {\it Hipparcos}
non-detection favours a larger distance. We have adopted a distance of 165 pc. This
distance puts R Hya just beyond the edge of the Local Bubble in this
direction \citep{lallement03}. At
our chosen distance, the proper motion is equivalent to a transverse
velocity of 48.5 \kms, and a space velocity of 49.5 \kms. This is within
$2\sigma$ of the average space velocity for AGB stars, $30$ \kms\
\citep{feast00}.

\begin{figure*}
\begin{center}
\includegraphics[angle=0,width=14cm]{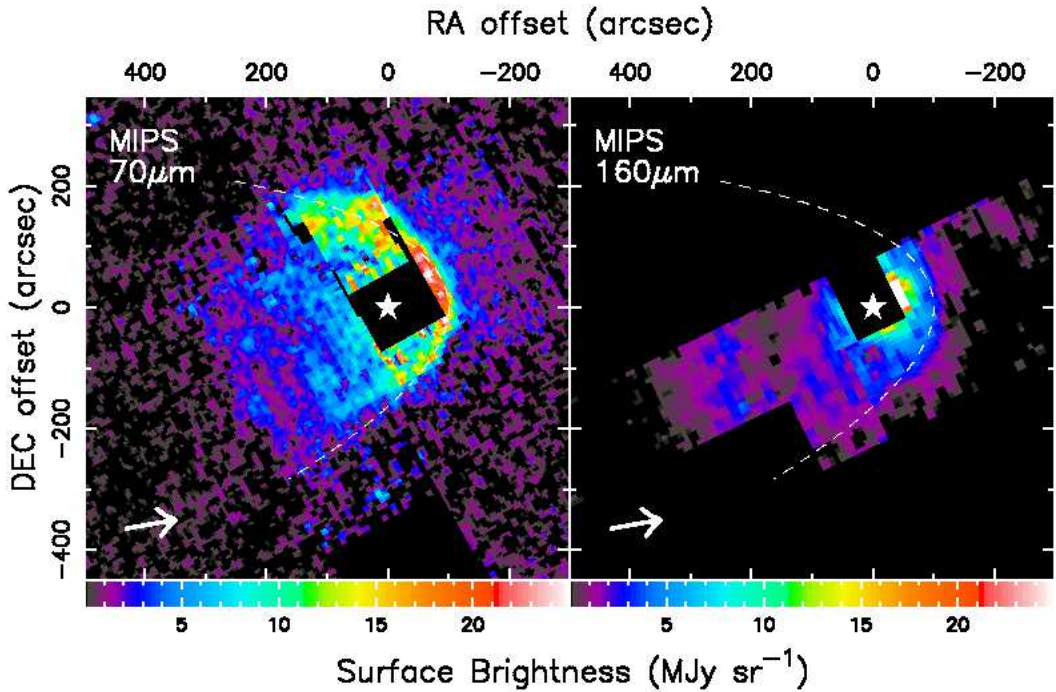} 
\caption{The background-subtracted, mosaicked MIPS colour maps of R Hya
at 70 $\mu$m (left) and 160 $\mu$m (right) \protect\citep{ueta06}. The images 
are zeroed at the position of the star (indicated by the `star') with North up
and East to the left. The proper motion is indicated by the arrow in the bottom
left. The parabolic arc is marked for clarity by the dashed line. Colour-scaling 
of surface brightness (in Mjy\,sr$^{-1}$) is provided by the colour bar.}
\label{MIRIAD}
\end{center}
\end{figure*}

High angular resolution observations were taken with the {\it Spitzer}
Space Telescope \citep{werner04} as part of the MIRIAD project
\citep{ueta06}. The Multiband Imaging Photometer aboard {\it Spitzer} (MIPS;
Rieke et al. 2004) image is shown in Fig. \ref{MIRIAD}. The central
unobserved region in the images avoids the central star which 
would saturate the detectors. For
details of the data reduction process, please consult \cite{ueta06}.

These images show that the material around the central star is not a
circular shell as suggested by the IRAS observations \citep{hashimoto98},
but instead
consists of a parabolic arc to the West of the star and regions of
emission to the East.  The arc appears brightest in the direction of
the proper motion, fading as it stretches away North and South. At its
brightest point, the arc is 100 arcsec from the central star. 
The arc is 300
arcsec wide across the star. There appears to be no emission
further West of the arc, suggesting there is no material related to R
Hya West of the arc. It is not clear from these images that the arc is
detached from the star as the region between the arc and star is in
the central unobserved region. To the East of the star, several
low-brightness regions of emission are present up to 300 arcsec
from the star. The location and brightness of these regions would
suggest they are associated with the star. Our simulations closely 
reproduce the morphology of the arc and the Eastern emission.

\section{Simulations}

Our simulations have been performed using a parallelised computational fluid
dynamics program designed to solve the standard Euler equations
of hydrodynamics using a second-order Godunov scheme due to
\cite{falle91}. It is in three dimensions
using Cartesian coordinates and includes the effect of radiative
cooling above $10^4$ K via a parametric fit to the cooling curve of
\cite{raymond76}. The timestep is defined by the
Courant-Friedrichs-Lewy condition with a Courant number of 0.5. We
also include a numerical viscosity to avoid on-axis issues in the
simulations.

Using this numerical scheme, we have modelled the interaction of a
stellar wind ejected from a star as it moves through the ISM. The
simulation is performed in the frame of reference of the star and the
motion through the ISM is like that of an oncoming wind - hence our
`two-wind' model.  
Further details of the model are explained in \cite{wareing06}.
We model a period of $5.5\times10^4$ years on the AGB and have
used a simulation grid of $200^3$ cubic cells (cell size of
$4\times10^{-3}$pc) producing a grid of 0.8 pc in each direction.

We set the AGB wind parameters of the model with a mass-loss rate of
$3 \times 10^{-7}$ \mdot\ and a velocity of 10 \kms\ 
(Zijlstra et al. 2002).
The temperature of the wind is set at 10$^4$ K, which is
the lowest value for which the cooling function is defined
\citep{wareing05}. The real temperature will be considerably less than
this and indeed the temperature of the undisturbed AGB wind in the 
simulation is on the order of a few tens of K.
Our simulations do not model dust physics or radiation
transport.  The gas pressure in both winds is calculated assuming an
ideal gas equation of state with an adiabatic index of $5/3$ in both
winds, effectively ignoring molecules.  

The position of the bow shock 0.08 pc ahead of the star at our 
adopted distance can be understood in
terms of a ram pressure balance. This balance predicts a local ISM
density of 0.6 H\,cm$^{-3}$ which we take as the ISM density in our
simulations.  The ISM is modelled as a warm neutral medium with a
temperature of $8 \times 10^3$ K. The stellar motion corresponds to 
a Mach number of 3.65 and the AGB wind has a Mach number of 0.74.
At 165 pc, R Hya is located 101 pc above the Galactic plane. ISM density, and
hence pressure, is constant at $n_{\rm H} = 2$ cm$^{-3}$
\citep{spitzer78} up to a scale height of 100 pc \citep{binney98}. 
Above this, density drops off exponentially, giving
an ISM density in the region of R Hya of 0.74 H\,cm$^{-3}$ 
in accidental agreement with our simulation. This is also
consistent with the estimate of 0.4 H\,cm$^{-3}$ calculated by \cite{ueta06}.
Densities within the Local Bubble would be very much lower: the
bow shock supports a location of R Hya beyond this region. 
 
\section{Results}

\begin{figure}
\begin{center}
\includegraphics[angle=0,width=7cm]{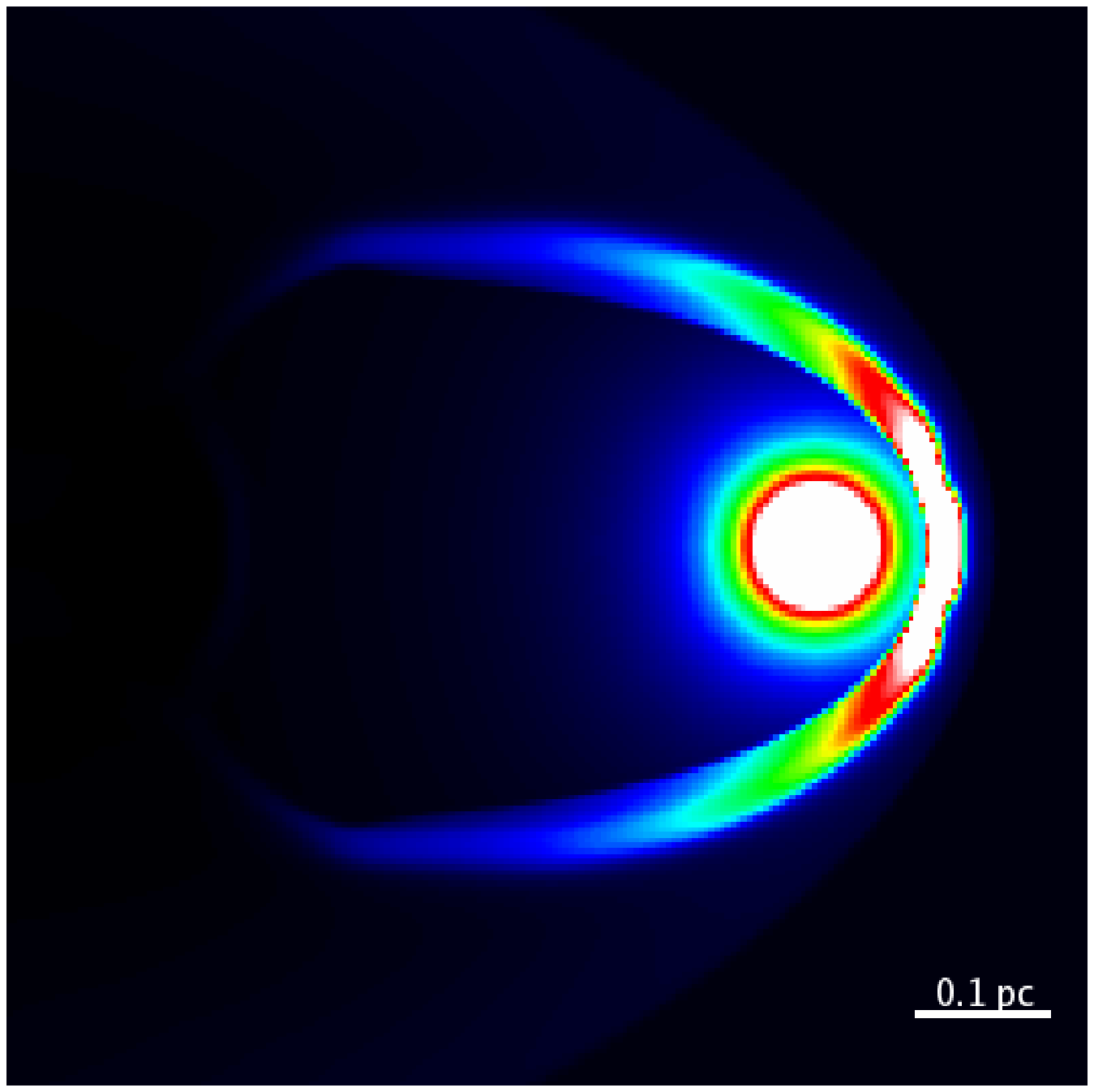} \\
\includegraphics[angle=0,width=7cm]{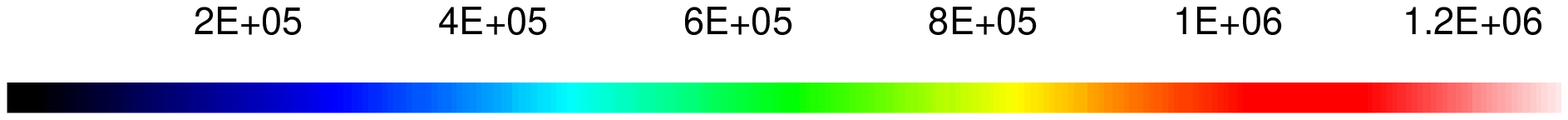} 
\caption{An image showing the gas density in the plane in which the 
star is located, parallel to the direction of motion, 40\,000 years 
into the simulation. The ISM is flowing in from the right. 
This image is at 9.5$^\circ$ to the actual line of
sight to R Hya.  The colour bar is
in units of \density\ where $2.4 \times 10^5$ corresponds to 10 
H\,cm$^{-3}$. The image is 0.8 pc on a side, which corresponds to
the physical image scale in Fig. \ref{MIRIAD} at 165 pc.}
\label{simulation}
\end{center}
\end{figure}

Fig. \ref{simulation} shows density on the plane in which the star is located
40\,000 years into the AGB phase. This point in time is
an arbitrary choice since the wind structure reached a stable
state after 25\,000 years into the simulation. The image is at 
$9.5^\circ$ to the actual line of sight to R Hya.

The wind from the star has driven a shock into the ISM. In the case of
a stationary star, this is a spherical shell of AGB wind
material sweeping up shocked ISM material. In the moving case, the
shock forms into a bow shock expanding ahead of the
star. Eventually, the bow shock reaches a maximum distance ahead of the
star (after 25\,000 years in our simulation) which can be understood
in terms of a ram pressure balance. Material is being
ram-pressure-stripped from the head of the bow shock into a tail
stretching downstream. The `bow shock and tail' structure is the
same as we have seen in other simulations considering a range of
space velocities
and wind parameters. We note this as support for the formation of this
general structure as a convergent result.

The width
of the bow shock across the star is approximately 0.24 pc agreeing with
the observations. If the bow shock is a strong shock, the temperature
of the shocked material at the head can be predicted by T $\sim 3/16\
m\,v^2/k\ =\ 33\,000$ K, where $m$ is taken as the average gas
component mass of $1.0\times10^{-27}$ kg, $v$ is the speed of the
central star relative to the ISM and $k$ is the Boltzmann
constant. The temperature in the simulation at the head of the bow
shock is found to be in agreement with this prediction. The
survival of dust above temperatures of approximately 1500 K 
is questionable, but at the low
densities in the simulation, the dust and gas are decoupled and
evidently the dust temperature can be much lower.

\section{Discussion}

\subsection{The bow shock}

The two-wind model can reproduce the appearance of the circumstellar
structure around R Hya, with physical dimensions matching that of the
bow shock. The head of the bow shock is in good agreement
with the direction of motion. Small regions of emission downstream
can be explained in terms of regions of higher density ISM
encountered by the stellar wind. Such high density regions in the ISM
have been shown to survive ablation by stellar winds \citep{pittard05}.


In our simulation, we find a high temperature of 35\,000 K at the head
of the bow shock, with material cooling rapidly as it moves down the
tail. The high temperatures are consistent with the observed H$\alpha$
emission \citep{gaustad01,ueta06}. The high temperatures will affect the dust 
by collisional heating. However,  it is impossible for us to
say whether this heat from the local gas is more important than the
stellar radiation or the interstellar radiation field 
(Speck et al. 2000)
in controlling the dust temperature.

The {\it Spitzer} and {\it IRAS} images indicate that the bow shock region is
detected through dust emission; the fact that the emission is strongest
at 60$\mu$m indicates a dust temperature of order 60 K. But other
emission processes may play a role. Shocked regions can show
[O\,{\sc{i}}] 63$\mu$m and 146$\mu$m lines. The {\it IRAS} 100$\mu$m detection
\citep{hashimoto98}, in a band without strong lines, shows that dust
contributes to the far-IR emission. 

\subsection{Other cases: mass-loss variations versus ISM interaction}

Cases of detached shells around AGB stars include U Ant
\citep{izumiura97}, U Hya \citep{waters94}, Y CVn \citep{izumiura96},
R Hya \citep{hashimoto98} and IRAS02091+6333 \citep{zijlstra02a}
observed in the far-infrared; detached molecular (CO) gas shells
around TT Cyg, S Sct, R Scl, U Ant, U Cam, V644 Sco and DR Ser
(Olofsson et al. 1996, Sch{\"o}ier et al. 2005).
All are carbon stars, apart from
R Hya and IRAS02091+6333. It has been suggested that the thermal pulse
scenario may only lead to mass-loss spikes for carbon
stars (Sch{\"o}ier et al. 2005)
but this is very much dependent on the assumed
mass loss prescription which is not well known. {\it IRAS} colours indicate that
detached shells do also exist around oxygen-rich stars \citep{zijlstra92}
and many more stars are likely to show large shells 
(Young et al. 1993).

An interesting comparison can be made between R Hya and U Hya. U Hya
has a thin circular shell with a radius of 120 arcsec
\citep{waters94}. It has a space velocity of 50 \kms\ and the parallax
gives a distance of 162 pc \citep{perryman97}.  The circular nature
of the shell suggests an internal origin such as the mass-loss
variations proposed by, for example, \cite{zijlstra92} and 
Sch{\"o}ier et al. (2005).
The age of the shell in this particular case is 6000 years.
The ISM interaction may be radially further away from
the star.


The star TT Cyg is surrounded by a circular detached shell at a radius
of approximately 40 arcsec \citep{olofsson96}; the physical size
of the shell is very similar to U Hya, in view of the larger distance
of TT Cyg. The thin symmetrical appearance supports an internal
origin. Interestingly, the space motion of TT Cyg (50 \kms) is almost
all in the radial direction away from us and any bow shock 
formed by an interaction with the ISM would appear circular on the
sky. But the fact that the slight offset of the star from the centre
of the shell is at right angles to the direction of proper motion,
favours an internal origin in mass-loss variations.

For other stars there is insufficient data to decide on the cause of
the detached shells.  It may be that the well-studied carbon stars with
detached shells are mostly due to internal mechanisms, i.e. thermal
pulses, while the fainter, less studied shells are dominated by ISM
interactions. CO observations of shells around carbon stars 
reveal relatively high velocities (about 20 \kms) which favours
thermal pulse origins (Olofsson et al. 1996, Sch{\"o}ier et al. 2005).  
Multiple detached shells do exist in the AGB phase of
evolution, e.g. S Sct and U Ant \citep{gonzalez03} and also R Hya:
thus, both mechanisms may occur simultaneously.

Young et al. (1993) in their analysis of 76 AGB stars resolved by {\it IRAS}
found no evidence for distortions by interaction with the ISM. Since R Hya 
is included in their sample, this lack of evidence
can be attributed to the poor spatial resolution and image quality 
of the {\it IRAS} instrument. 

R Hya can be considered a typical case for mass-loss rate and ISM
density although its space velocity is perhaps high.  Lower velocity
objects will have a bow shock and with higher mass-loss rates this
will be located further from the star. In a zero velocity case, the
bow shock transforms into a spherical swept-up shell of ISM material
(Young et al. 1993, Speck et al. 2000, Zijlstra \& Weinberger 2002). 
We predict that all AGB stars will show
some degree of an AGB-ISM interaction, although bow shocks will be
rarer. The interpretation of a detached shell in terms of a mass-loss
variation must be considered with this ISM interaction in mind.

\subsection{Mass loss history}

This model has shown that information usually gleaned from
circumstellar dust shells around Mira variables can no longer be
inferred in this situation. The bow shock has destroyed any mass loss
history older than about $10^4$ years in the case of R Hya. Higher
mass-loss rate and/or lower lower ISM density could increase this
timescale as found by Young et al. (1993) and Zijlstra \& Weinberger (2002).

\begin{figure}
\begin{center}
\includegraphics[width=7cm]{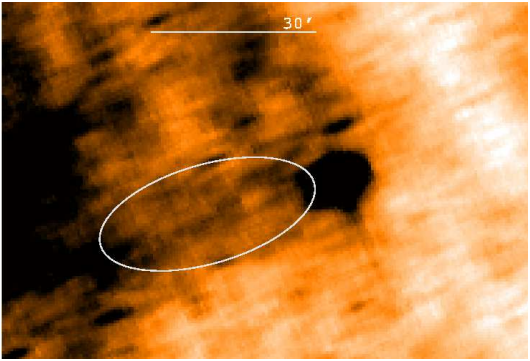}
\caption{A figure showing an {\it IRAS} $60 \mu$m observation of the area around
R Hya. North is up and East
to the left. Evidence for a tail of material is indicated by the ellipse.}
\label{iras}
\end{center}
\end{figure}
 
The simulations predict the occurence of a tail, consisting of
swept-back ISM and stellar wind gas. The mass loss history can in
principle still be traced down the length of the tail. {\it IRAS} maps provide
some indication for a tenuous detection of material downstream of R
Hya as shown in Fig. \ref{iras}. This material stretches up to 30
arcmin away or 1.5 pc at a distance of 165 pc. At 49.5 \kms\ this
implies a minimum tail age of 30\,000 years. Adding the 25\,000 years
it takes to form the stable bow shock, to represent the travel time from the
star to the bow shock and down the tails, we predict R Hya has been
losing mass for at least 55\,000 years.

If we consider it has taken 25\,000 years to form the bow shock and
after this the bow shock is in a steady state, an appreciable amount
of mass is in the bow shock. The estimate from the simulation is 
$2.6\times10^{-3}$ \msun\ in the region of the bow shock defined as a
hemispherical upwind shell centred on the star with inner radius of 0.08
pc and a thickness of 0.1 pc. Our analytical estimate of the mass in 
the bow shock is consistent with this estimate, suggesting there is four
times more stellar mass in the bow shock than ISM mass. At a dust 
temperature of 60 K and assuming a dust to gas ratio of 100, we predict
a 100 $\mu$m flux of 15 Jy, in broad agreement with the fluxes observed by 
\cite{ueta06}.

We use a constant stellar mass-loss rate and wind velocity. However,
the dust shell models of various people 
(\protect{Zijlstra} \& Weinberger 1992, Sch{\"o}ier et al. 2005)
indicate that these quantities are variable, particularly during the
thermal pulses. Models indicate that in the
course of thermal pulses the stellar radius changes temporarily by a
factor of 2, and the mass-loss rate by an order of magnitude, causing 
a density and velocity spike moving radially away from the star - considered as
the origin of detached dust shells. Such events could severely disturb
the pressure balance that dictates the well defined location of the
bow shock when one of these `wind'-shells runs into the bow
shock. There is likely to be a complex time dependence of the wind-ISM
interaction on the timescales of the thermal pulses, as has been
considered in 2D simulations by 
Villaver et al. (2003).

\section*{Acknowledgments}

The numerical computations in this work were carried out using the
COBRA supercomputer at Jodrell Bank Observatory. The {\it
Spitzer} Space Telescope is operated by the JPL/Caltech under a contract
with NASA. We acknowledge additional support for the following
individuals: Speck - NASA ADP grant (NAG 5-12675), Ueta - 
NPP Research Fellowship Award, Elitzur - NSF Grant: AST--0507421, Gehrz - in part
by NASA (Contract 1215746) issued by JPL/Caltech to Gehrz to the
University of Minnesota, Herwig - LDRD program (20060357ER)
at LANL, Izumiura - Grant-in-Aid (C) from JSPS (No.17540221), 
Matsuura - JSPS, Stencel - in part by NASA (Contract 1275979)
issued by JPL/Caltech to the University of Denver, Szczerba - 
grant 2.P03D.017.25, Wareing - PPARC.

\label{lastpage}


\begin{thebibliography}{99}

\bibitem[\protect\citeauthoryear{Binney \& Merrifield}{1998}]{binney98} Binney J., Merrifield M., 1998, in Galactic Astronomy (Princeton, NJ: Princeton University Press), Ch. 10
\bibitem[Eggen(1985)]{eggen1985} Eggen, O.J.\ 1985, AJ, 90, 333 
\bibitem[\protect\citeauthoryear{Falle}{1991}]{falle91} Falle S.A.E.G., 1991, MNRAS, 250, 581
\bibitem[\protect\citeauthoryear{Feast \& Whitelock}{2000}]{feast00} Feast M.W., Whitelock P.A, 2000, MNRAS, 317, 460
\bibitem[\protect\citeauthoryear{Gaustad et al.}{2001}]{gaustad01} Gaustad J.E., McCullough P.R., Rosing W., Van Buren D., 2001, PASP, 113, 1326
\bibitem[\protect\citeauthoryear{Gonz{\'a}lez\ Delgado et al.}{2001}]{gonzalez01} Gonz{\'a}lez\ Delgado D., Olofsson H., Schwarz H.E., Eriksson K., Gustafsson B., 2001, A\&A, 372, 885 
\bibitem[\protect\citeauthoryear{Gonz{\'a}lez\ Delgado et al.}{2003}]{gonzalez03} Gonz{\'a}lez\ Delgado D., Olofsson H., Schwarz H.E., Eriksson K., Gustafsson B., Gledhill T., 2003, A\&A, 399, 1021
\bibitem[\protect\citeauthoryear{Hashimoto et al.}{1998}]{hashimoto98} Hashimoto O., Izumiura H., Kester D.J.M., Bontekoe T.R., 1998, A\&A, 329, 213
\bibitem[\protect\citeauthoryear{Izumiura et al.}{1996}]{izumiura96} Izumiura H., Hashimoto O., Kawara K., Yamamura I., Waters L.B.F.M., 1996, A\&A, 315, L221 
\bibitem[\protect\citeauthoryear{Izumiura et al.}{1997}]{izumiura97} Izumiura H. et al. 1997, A\&A, 323, 449
\bibitem[\protect\citeauthoryear{Lallement et al.}{2003}]{lallement03} Lallement R., Welsh B.Y., Vergely J.L., Crifo F., Sfeir D., 2003, A\&A, 411, 447 
\bibitem[\protect\citeauthoryear{Olofsson et al.}{1990}]{olofsson90} Olofsson H., Carlstrom U., Eriksson K., Gustafsson B., Willson L.A., 1990, A\&A, 230, L13
\bibitem[\protect\citeauthoryear{Olofsson et al.}{1996}]{olofsson96} Olofsson H., Bergman P., Eriksson K., Gustafsson B., 1996, A\&A, 311, 587
\bibitem[\protect\citeauthoryear{Perryman et al.}{1997}]{perryman97} Perryman M.A.C. et al., 1997, A\&A, 323, L49
\bibitem[\protect\citeauthoryear{Pittard et al.}{2005}]{pittard05} Pittard J.M., Dyson J.E., Falle S.A.E.G., Hartquist T.W., 2005, MNRAS, 361, 1077
\bibitem[\protect\citeauthoryear{Raymond, Cox \& Smith}{1976}]{raymond76} Raymond J.C., Cox D.P., Smith B.W., 1976, ApJ, 204, 290
\bibitem[\protect\citeauthoryear{Rieke et al.}{2004}]{rieke04} Rieke G.H. et al. 2004, ApJS, 154, 25
\bibitem[\protect\citeauthoryear{Sch{\"o}ier, Lindqvist \& Olofsson}{2005}]{schoier05} Sch{\"o}ier F.L., Lindqvist M., Olofsson H., 2005, A\&A, 436, 633
\bibitem[\protect\citeauthoryear{Speck, Meixner \& Knapp}{2000}]{speck00} Speck A.K., Meixner M., Knapp G.R., 2000, ApJ, 545, L145
\bibitem[\protect\citeauthoryear{Spitzer}{1978}]{spitzer78} Spitzer L. Jr, 1978, Physical Processes in the Interstellar Medium (New York: Wiley), p. 234
\bibitem[\protect\citeauthoryear{Steffen \& Sch{\"o}nberner}{2000}]{steffen00} Steffen M., Sch{\"o}nberner D., 2000, A\&A, 357, 180
\bibitem[\protect\citeauthoryear{Ueta et al.}{2006}]{ueta06} Ueta T. et al. 2006, ApJL, {\it submitted}
\bibitem[\protect\citeauthoryear{Vassiliadis \& Wood}{1993}]{vassiliadis93} Vassiliadis E., Wood P., 1993, ApJ, 413, 641
\bibitem[\protect\citeauthoryear{Villaver, Garcia-Segura \& Manchado}{2003}]{villaver03} Villaver E., Garcia-Segura G., Manchado A., 2003, ApJL, 585, L49
\bibitem[\protect\citeauthoryear{Waters et al.}{1994}]{waters94} Waters L.B.F.M., Loup C., Kester D.J.M., Bontekoe T.R., de Jong T., 1994, A\&A, 281, L1
\bibitem[\protect\citeauthoryear{Wareing}{2005}]{wareing05} Wareing C.J. 2005, PhD thesis, Univ. of Manchester, 2005
\bibitem[\protect\citeauthoryear{Wareing et al.}{2006}]{wareing06} Wareing C.J., O'Brien T.J., Zijlstra A.A., Kwitter K.B., Irwin J., Wright N., Greimel R., Drew J., 2006, MNRAS, 366, 387
\bibitem[\protect\citeauthoryear{Werner et al.}{2004}]{werner04} Werner M.W. et al. 2004, ApJS, 154, 1
\bibitem[\protect\citeauthoryear{Wilson}{1953}]{wilson53} Wilson R.E. 1953, General Catalogue of Stellar Radial Velocities (Carnegie Institute Washington D.C. Publication)
\bibitem[\protect\citeauthoryear{Young, Phillips \& Knapp}{1993}]{young93} Young K., Phillips T.G., Knapp G.R., 1993, ApJ, 409, 725 
\bibitem[\protect\citeauthoryear{Zijlstra et al.}{1992}]{zijlstra92} Zijlstra A.A., Loup C., Waters L.B.F.M., de Jong T., 1992, A\&A, 265, L5
\bibitem[\protect\citeauthoryear{Zijlstra \& Weinberger}{2002}]{zijlstra02a} Zijlstra A.A., Weinberger R., 2002, ApJ, 572, 1006
\bibitem[\protect\citeauthoryear{Zijlstra, Bedding \& Mattei}{2002}]{zijlstra02b} Zijlstra A.A., Bedding T.R, Mattei J.A., 2002, MNRAS, 334, 498

\end{thebibliography}
\end{document}